\documentclass[fleqn,usenatbib,usedcolumn]{mnras}

\usepackage[british]{babel}

\usepackage{ae,aecompl}
\usepackage[T1]{fontenc}
\usepackage{booktabs}

\usepackage{graphicx}
\usepackage{amsmath}
\usepackage{amssymb}

\newcommand{\bagpipes}{\textsc{Bagpipes}}
\newcommand*\samethanks[1][\value{footnote}]{\footnotemark[#1]}

\defcitealias{Belli2019}{B19}

\title[Massive quiescent galaxies at $2 < z < 5$]{Timing the earliest quenching events with a robust sample of massive quiescent galaxies at 2\,<\,z\,<\,5}

\author[A. C. Carnall et al.]{A. C. Carnall$^{1}$\thanks{E-mail: adamc@roe.ac.uk},
S. Walker$^{1}$,
R. J. McLure$^{1}$,
J. S. Dunlop$^{1}$,
D. J. McLeod$^{1}$,
\newauthor F. Cullen$^{1}$,
V. Wild$^{2}$,
R. Amorin$^{3, 4}$,
M. Bolzonella$^{5}$,
M. Castellano$^{6}$,
A. Cimatti$^{7, 8}$,
\newauthor O. Cucciati$^{5}$,
A. Fontana$^{6}$,
A. Gargiulo$^{9}$,
B. Garilli$^{9}$,
M. J. Jarvis$^{10, 11}$,
L. Pentericci$^{6}$,
\newauthor L. Pozzetti$^{5}$,
G. Zamorani$^{5}$,
A. Calabro$^{6}$,
N. P. Hathi$^{12}$,
A. M. Koekemoer$^{12}$
\\
\\
$^{1}$ SUPA\thanks{Scottish Universities Physics Alliance}, Institute for Astronomy, University of Edinburgh, Royal Observatory, Edinburgh EH9 3HJ, UK\\
$^{2}$ SUPA\samethanks, School of Physics and Astronomy, University of St. Andrews, North Haugh, St. Andrews KY16 9SS, UK \\
$^{3}$ Instituto de Investigaci\'on Multidisciplinar en Ciencia y Tecnolog\'ia, Universidad de La Serena, Ra\'ul Bitr\'an 1305, La Serena, Chile \\
$^{4}$ Departamento de F\'isica y Astronom\'ia, Universidad de La Serena, Av. Juan Cisternas 1200 Norte, La Serena, Chile \\
$^{5}$ INAF - OAS Bologna, Via P. Gobetti 93/3, I-40129, Bologna, Italy \\
$^{6}$ INAF - Osservatorio Astronomico di Roma, Via Frascati 33, I-00078 Monteporzio Catone, Italy \\
$^{7}$ University of Bologna, Department of Physics and Astronomy (DIFA), Via Gobetti 93/2, I-40129, Bologna, Italy \\
$^{8}$ INAF - Osservatorio Astrofisico di Arcetri, Largo E. Fermi 5, I-50125, Firenze, Italy \\
$^{9}$ INAF - IASF Milano, Via A. Corti 12, I-20133, Milano, Italy \\
$^{10}$ Department of Physics, University of Oxford, Keble Road, Oxford OX1 3RH, UK \\
$^{11}$ Department of Physics \& Astronomy, University of the Western Cape, Private Bag X17, Bellville, Cape Town, 7535, South Africa \\
$^{12}$ Space Telescope Science Institute, 3700 San Martin Dr., 
Baltimore, MD 21218, USA
}
\date{Accepted XXX. Received YYY; in original form ZZZ}
\pubyear{2019}

\begin{document}
\label{firstpage}
\pagerange{\pageref{firstpage}--\pageref{lastpage}}
\maketitle

\begin{abstract}

\noindent We present a sample of 151 massive ($M_* > 10^{10}\mathrm{M_\odot}$) quiescent galaxies at ${2 < z < 5}$, based on a sophisticated Bayesian spectral energy distribution fitting analysis of the CANDELS UDS and GOODS-South fields. Our sample includes a robust sub-sample of 61 objects for which we confidently exclude low-redshift and star-forming solutions. We identify 10 robust objects at $z>3$, of which 2 are at $z>4$. We report formation redshifts, demonstrating that the oldest objects formed at $z > 6$, however individual ages from our photometric data have significant uncertainties, typically $\sim0.5$ Gyr. We demonstrate that the UVJ colours of the quiescent population evolve with redshift at $z>3$, becoming bluer and more similar to post-starburst galaxies at lower redshift. Based upon this we construct a model for the time-evolution of quiescent galaxy UVJ colours, concluding that the oldest objects are consistent with forming the bulk of their stellar mass at $z\sim6-7$ and quenching at $z\sim5$. We report spectroscopic redshifts for two of our objects at $z=3.440$ and $3.396$, which exhibit extremely weak Ly$\alpha$ emission in ultra-deep VANDELS spectra. We calculate star-formation rates based on these line fluxes, finding that these galaxies are consistent with our quiescent selection criteria, provided their Ly$\alpha$ escape fractions  are $>3$ and $>10$ per cent respectively. We finally report that our highest redshift robust object exhibits a continuum break at $\lambda\sim7000$\AA\ in a spectrum from VUDS, consistent with our photometric redshift of $z_\mathrm{phot}=4.72^{+0.06}_{-0.04}$. If confirmed as quiescent, this object would be the highest redshift known quiescent galaxy. To obtain stronger constraints on the times of the earliest quenching events, high-SNR spectroscopy must be extended to $z\gtrsim3$ quiescent objects.
\end{abstract}
\clearpage

\section{Introduction}\label{sect:intro}

Understanding the processes that shaped the most massive galaxies in the local Universe is one of the major challenges in modern astrophysics. Historically the age-metallicity-dust degeneracy that plagues photometric studies (e.g. \citealt{Conroy2013}) has limited our ability to extract detailed physical information from galaxy spectral energy distributions (SEDs). However since the advent of large spectroscopic surveys it has been firmly established that local massive galaxies exhibit little ongoing star-formation activity, possessing old, passively evolving stellar populations (e.g. \citealt{Kauffmann2003, Brinchmann2004, Gallazzi2005}).

Even with spectroscopic data it is still challenging to determine the redshift at which the bulk of the stellar mass in local quiescent galaxies formed, owing to the extremely slow time-evolution of old stellar populations. Instead, as instrumental capabilities have expanded, efforts have been made to learn about the formation histories of these objects by studying their assumed progenitors, in the form of high-redshift massive galaxies. These are now routinely studied in detail during their star-forming phases, both in the rest-frame UV-optical (e.g. \citealt{Kriek2015}; \citealt{McLure2018a}; \citealt{Cullen2019}) and far-infrared (e.g. \citealt{Dunlop2017}; \citealt{Stach2019}; \citealt{Williams2019}). However once star-formation ceases galaxies rapidly become fainter and redder, making the extension of spectroscopic studies to quiescent objects at the highest redshifts challenging.

Spectroscopic studies at $z\sim1{-}2$ have long demonstrated that old, passively evolving galaxies already exist at this epoch (e.g. \citealt{Dunlop1996}; \citealt{Cimatti2004, Cimatti2008}; \citealt{Daddi2005}; \citealt{Whitaker2013}) and recent work has extended these studies to robust statistical samples (e.g. \citealt{Belli2019}; \citealt{Carnall2019b}). This finding suggests a gap in our knowledge between the most massive star-forming galaxies at $z\gtrsim3$ and already-old quiescent galaxies at $z\lesssim2$. In order to understand this transition in detail, it is imperative that large samples of spectroscopic data are collected for objects at $z>2$ that show signs of being the first galaxies in the Universe to quench their star-formation activity. Such studies would allow the precise characterisation of the onset times and time-scales of quenching for old massive quiescent galaxies at $z\lesssim2$, and also provide constraints on the growth of such systems through mergers post-quenching.

Recently, preliminary spectroscopic studies at $z>3$ have begun to confirm the existence of quiescent objects at these redshifts (e.g. \citealt{Glazebrook2017}; \citealt{Schreiber2018}; \citealt{Valentino2019}; \citealt{Forrest2019}), for which strong photometric evidence has been available for several years (e.g. \citealt{Straatman2014, Straatman2016}; \citealt{Nayyeri2014}). These studies involve attempts to observe Balmer absorption features in the near-infrared $H$ and $K$-bands, however the limitations of existing instrumentation mean that such studies must be confined to the most extreme objects, and cannot reach the depths required to place strong constraints on physical properties such as stellar ages and metallicities. 

As we move into the 2020s new instruments, such as NIRSpec on-board the \textit{James Webb Space Telescope} and HARMONI at the Extremely Large Telescope, will provide the ability to observe the highest redshift quiescent galaxies with sufficient signal-to-noise ratio (SNR) to extract subtle physical properties. This will afford us a new view of the ultimate origins of local massive galaxies, as well as a deeper understanding of the quiescent galaxy population at high redshift, which is currently poorly reproduced by cosmological simulations (e.g. \citealt{Dave2017}; \citealt{Cecchi2019}).

In this work we seek to define a robust and representative sample of massive quiescent galaxies at $z > 2$ from the extremely high quality photometric data available in CANDELS UDS and GOODS-South (see Section \ref{sect:method_data}). This analysis will allow us to set preliminary constraints on the number densities and physical properties of these objects, and serve as a basis for large, detailed spectroscopic follow-up programmes with current and future instruments.

Whilst similar attempts have been made for many years (e.g. \citealt{Chen2004}; \citealt{Mancini2009}; \citealt{Fontana2009}) several significant challenges remain. Firstly, the age-metallicity-dust degeneracy leads to significant uncertainties in the samples selected, with contemporary studies still experiencing contamination at levels of $\gtrsim20$ per cent (\citealt{Schreiber2018}). A specific issue is the presence of strong emission lines in high-redshift dusty star-forming galaxies, in particular [O\,\textsc{iii}] at 5007\AA, which can mimic the near-infrared colours associated with the strong Balmer breaks of recently quenched galaxies (e.g. \citealt{Merlin2018}).

In addition, it is also known that the specifics of the modelling assumptions made when performing SED fitting significantly impact the physical parameter values obtained (e.g. \citealt{Carnall2019a}; \citealt{Leja2019a}), with many analyses basing their conclusions on codes optimised for photometric redshift determination rather than detailed physical parameter recovery. Finally, there is no consensus on the optimal method for separating star-forming and quiescent galaxies, particularly at high redshift, with a variety of colour and specific star-formation rate (sSFR) criteria in use (e.g. \citealt{Williams2009}; \citealt{Fontana2009}; \citealt{Pacifici2016}; \citealt{Merlin2018}; \citealt{Leja2019c}; \citealt{Girelli2019}). 

In this work we use Bayesian Analysis of Galaxies for Physical Inference and Parameter EStimation (\bagpipes; \citealt{Carnall2018}) to address these issues. \bagpipes\ is one of a new generation of spectral modelling and fitting tools, which combines highly sophisticated physical models for galaxy spectra with a fully Bayesian fitting approach. One of the major advantages of \bagpipes\ is that the observed redshift can be fitted in parallel with sophisticated physical models. This, combined with the implementation of the \textsc{MultiNest} nested sampling algorithm (\citealt{Feroz2008, Feroz2009, Feroz2013}), allows detailed recovery of full posterior distributions for physical parameters, including clear identification of secondary low-redshift or star-forming solutions for candidate high-redshift quiescent galaxies.

We begin in Section \ref{sect:method} by introducing the data, fitting methodology and selection criteria we employ to obtain our sample. In Section \ref{sect:results} we report inferred UVJ colours, formation redshifts and number densities. In Section \ref{sect:discussion:oldest} we present an analysis designed to understand when the oldest galaxies in our sample quenched their star formation. In Section \ref{sect:discussion:z3} we focus on our individual robustly identified objects at $z>3$, reporting probable spectroscopic redshifts for 3 objects. We present our conclusions in Section \ref{sect:conclusion}. All magnitudes are quoted in the AB system. For cosmological calculations we adopt $\Omega_M = 0.3$, $\Omega_\Lambda = 0.7$ and $H_0$ = 70 $\mathrm{km\ s^{-1}\ Mpc^{-1}}$. All times, $t$, are measured forwards from the beginning of the Universe. We assume a \cite{Kroupa2001} initial mass function.

\begin{table*}
  \caption{Parameters and priors for the model we fit to our data (see Section \ref{sect:method_fitting}). For Gaussian priors, $\mu$ is the mean and $\sigma$ the standard deviation of the prior distribution. The upper limit on the $\tau$ parameter, $t_\mathrm{obs}$, is the age of the Universe at redshift $z$. Logarithmic priors are uniform in log base ten of the parameter. We use a double power law SFH model and the dust attenuation law of \protect \cite{Salim2018}.}
\begingroup
\setlength{\tabcolsep}{10pt} 
\renewcommand{\arraystretch}{1.1} 
\begin{tabular}{llllll}
\hline
Parameter & Symbol / Unit & Range & Prior & \multicolumn{2}{l}{Hyper-parameters} \\
\hline
Redshift & $z$ & (0, 10) & Uniform & & \\
Stellar mass formed & $M_*\ /\ \mathrm{M_\odot}$ & (1, $10^{13}$) &Logarithmic & & \\
Stellar metallicity & $Z\ /\ \mathrm{Z_\odot}$ & (0.1, 2.5) &Logarithmic & & \\
Double power law falling slope & $\alpha$ & (0.01, 1000) & Logarithmic & & \\
Double power law rising slope & $\beta$ & (0.01, 1000) & Logarithmic & & \\
Double power law turnover time & $\tau$ / Gyr & (0.1, $t_\mathrm{obs}$) & Uniform & & \\
$V$-band attenuation & $A_V$ / mag & (0, 8) & Uniform & & \\
Deviation from \cite{Calzetti2000} slope & $\delta$ & ($-0.3$, 0.3) & Gaussian & $\mu = 0$ & $\sigma$ = 0.1 \\
Strength of 2175\AA\ bump & $B$ & (0, 5) & Uniform & & \\
\hline
\end{tabular}
\endgroup
\label{table:params}
\end{table*}

\section{Method}\label{sect:method}

\subsection{Photometric Catalogues}\label{sect:method_data}

In this work we analyse data from CANDELS UDS and GOODS-South, two of the best-studied extragalactic legacy fields. Both have deep near-ultraviolet to near-infrared \textit{Hubble Space Telescope} (HST) data from CANDELS (\citealt{Grogin2011}; \citealt{Koekemoer2011}) and other programmes, as well as a wide range of ground-based and $Spitzer$-IRAC multi-wavelength ancillary data. Full details of the catalogues used can be found in \cite{Galametz2013} and \cite{Guo2013} for UDS and GOODS-S respectively.

These two fields in particular were chosen for this work due to the availability of ultra-deep $K$-band imaging from the Hawk-I UDS and GOODS Survey (HUGS; \citealt{Fontana2014}). At $z\gtrsim3$ the Balmer/4000\AA\ break falls between the $H$ and $K$ bands, meaning the extremely deep imaging in these bands provided by CANDELS and HUGS respectively is critical for the robust identification of older stellar populations at these redshifts. For this reason we supplement the \cite{Guo2013} GOODS-S catalogue with additional, deeper $K$-band data from the final HUGS data release.

Before fitting we select only objects with $H_{160} < 27$, leaving 61837 objects. We additionally remove all objects with missing observations in more than two bands ($\sim{}3$ per cent in UDS; $\sim{}5$ per cent in GOODS-S). When fitting we impose a maximum SNR (or error floor; e.g. \citealt{Muzzin2013}) to account for the error budget becoming dominated by systematic rather than random errors at high SNRs. We use a maximum SNR of 20 for all photometric bands except IRAC channels 1 and 2 for which we use a maximum SNR of 10 and IRAC channels 3 and 4 for which we use a maximum SNR of 5. The median SNR of the catalogue is 12.0 in the $H$-band, 5.0 in IRAC channel 1 and 1.2 in IRAC channel 3.

\begin{table*}
  \caption{Coordinates, magnitudes and \bagpipes\ physical parameters ($16^\mathrm{th}{-}84^\mathrm{th}$ percentiles) for our parent $z>2$ massive quiescent galaxy sample, selected in Section \ref{sect:method_selection}. The full table, including additional columns, is available as supplementary online material.}
\begingroup
\setlength{\tabcolsep}{10pt} 
\renewcommand{\arraystretch}{1.4} 
\begin{tabular}{lccccccc}
\hline
ID & RA & DEC & $F160W$ & $Ks$ & Redshift & $t_\mathrm{form}\ /\ $Gyr& log$_{10}(M_*$/M$_\odot)$ \\
\hline
GOODSS-00402&53.125301&-27.934856&23.4&23.2&$2.41^{+0.06}_{-0.05}$&$2.3^{+0.1}_{-0.1}$&$10.38^{+0.05}_{-0.04}$ \\
GOODSS-00954&53.161680&-27.918738&23.3&22.9&$2.54^{+0.05}_{-0.04}$&$2.0^{+0.1}_{-0.4}$&$10.42^{+0.05}_{-0.03}$ \\
GOODSS-01000&53.136832&-27.917348&24.5&23.6&$2.30^{+0.15}_{-0.15}$&$0.7^{+0.5}_{-0.4}$&$10.61^{+0.07}_{-0.06}$ \\
GOODSS-01086&53.137180&-27.915837&23.7&23.3&$2.70^{+0.05}_{-0.05}$&$2.0^{+0.1}_{-0.1}$&$10.26^{+0.05}_{-0.05}$ \\
GOODSS-01119&53.169052&-27.915846&22.0&21.6&$2.12^{+0.05}_{-0.05}$&$2.4^{+0.2}_{-0.7}$&$11.00^{+0.06}_{-0.04}$ \\
\hline
\end{tabular}
\endgroup
\label{table:sample}
\end{table*}

\subsection{Fitting with B\small{AGPIPES}}\label{sect:method_fitting}

Our main objective when designing the model to be fitted was to allow sufficient flexibility to permit all potential solutions over a wide range of observed redshifts and sSFRs. As such, we use a nine-parameter model including flexible prescriptions for galaxy star-formation histories (SFHs) and dust attenuation curve shapes. The parameters of our model are detailed below and summarised in Table \ref{table:params}.

We use the default \bagpipes\ stellar population models, which are the 2016 updated version of the \cite{Bruzual2003} models\footnote{\url{https://www.bruzual.org/~gbruzual/bc03/Updated_version_2016}} using the MILES stellar spectral library (\citealt{Falcon-Barroso2011}) and updated stellar evolutionary tracks of \cite{Bressan2012} and \cite{Marigo2013}. 

We parameterise the SFHs of our galaxies using the double power law form of \cite{Carnall2018}, which has been shown to reproduce well the SFHs of massive quiescent galaxies from the \textsc{Mufasa} simulation (\citealt{Dave2016}). We assume a single metallicity for all stars in our galaxies with scaled Solar abundances. This metallicity is allowed to vary with a uniform prior in logarithmic space over the range from $-1 < \mathrm{log}_{10}(Z_*/\mathrm{Z}_\odot) < 0.4$, where $\mathrm{Z}_\odot$ is defined as 0.02. The total stellar mass formed by each galaxy up to the observed redshift is fitted with a uniform prior in logarithmic space from $0 < \mathrm{log}_{10}(M_*/\mathrm{M}_\odot) < 13$.

We model dust attenuation in our galaxies using the flexible model of \cite{Salim2018}, which parameterises the dust curve shape in terms of a power-law deviation, $\delta$, from the \cite{Calzetti2000} law (see also \citealt{Noll2009}). As we are primarily interested in massive galaxies our prior is centred on the \cite{Calzetti2000} curve ($\delta=0$), which has been found to be a robust average attenuation curve shape for massive galaxies (e.g. \citealt{Cullen2017, Cullen2018}; \citealt{McLure2018a}). We set a Gaussian prior on $\delta$ with a standard deviation, $\sigma = 0.1$, and set a uniform prior on the 2175\AA\ bump strength, $B$, from 0 to 5, where the Milky Way law (e.g. \citealt{Cardelli1989}) has $B = 3$. We assume a uniform prior on $A_\mathrm{V}$ from 0 to 8 magnitudes. Whilst quiescent galaxies are the focus of our analysis, this wide range of $A_\mathrm{V}$ values is critical for correct identification of intermediate redshift ($z\sim1-3$) dusty star-forming galaxies, which, if misidentified, are known to contaminate high redshift quiescent galaxy samples (e.g. \citealt{Schreiber2018}). We further assume light from stars younger than 10 Myr and resulting nebular line and continuum emission experiences double the attenuation experienced by the rest of the stellar population.

Our model includes nebular line and continuum emission by post-processing of the stellar models using the \textsc{Cloudy} photoionization code (\citealt{Ferland2017}), following a method based on that of \cite{Byler2017}, as described in section 3.1.3 of \cite{Carnall2018}. The nebular metallicity is assumed to follow the stellar metallicity, and a fixed ionization parameter of log$_{10}(U) = -3$ is also assumed. 

We additionally include dust emission assuming energy balance between light attenuated and light re-radiated and using the dust emission models of \cite{Draine2007}. We assume values of 2 for $Q_\mathrm{pah}$, the percentage of dust mass in polycyclic aromatic hydrocarbons, 1 for $U_\mathrm{min}$, the minimum starlight intensity to which the dust is exposed, and 0.01 for $\gamma_\mathrm{e}$, the fraction of the incident starlight at $U_\mathrm{min}$. \bagpipes\ also includes attenuation due to the intergalactic medium using the model of \cite{Inoue2014}. We fit the observed redshift of each galaxy assuming a uniform prior.

\subsection{Selection of high-redshift quiescent galaxies}\label{sect:method_selection}

Having fitted the catalogues of Section \ref{sect:method_data} using the model of Section \ref{sect:method_fitting}, we calculate posterior percentiles for the physical parameters of each object, as well as minimum reduced chi-squared values, $\chi^2_\nu$. We then exclude objects with $\chi^2_\nu > 5.3$ in UDS and $\chi^2_\nu > 6.8$ in GOODS-S. These cuts correspond to $5\sigma$ upper limits on the expected $\chi^2_\nu$ values under the assumption that the fitted model correctly describes the data (the UDS catalogue has 19 bands, whereas the GOODS-S catalogue has 16, giving 9 and 6 degrees of freedom when fitting with our 9 parameter model). These cuts remove the worst-fitted $\sim{}13$ per cent of objects. The majority of these are contaminated due to close proximity with bright sources, whilst a minority have SEDs consistent with stars. We did not apply a point source selection due to concerns about wrongly excluding high redshift quiescent galaxies, which are known to be extremely compact (e.g. \citealt{McLure2013}). We then select objects with posterior median redshifts, $z > 2$, and stellar masses, $M_* > 10^{10}\mathrm{M_\odot}$.

The best method of separating star-forming and quiescent galaxies at any given epoch is still a subject of much debate. The two most common approaches are selection by rest-frame UVJ colours (e.g. \citealt{Williams2009}) and selection by sSFR (e.g. \citealt{Fontana2009}; \citealt{Gallazzi2014}; \citealt{Pacifici2016}; \citealt{Merlin2018}), with the selection criteria usually a function of observed redshift in both cases. In this work we use the most commonly applied sSFR selection criterion, requiring that galaxies have posterior median sSFR < 0.2/$t_\mathrm{H}$, where $t_\mathrm{H}$ is the age of the Universe at the posterior median observed redshift. In \cite{Carnall2018, Carnall2019b} we demonstrate good agreement between samples selected by this criterion and the \cite{Williams2009} $z=0$ UVJ criterion of $U - V$ > 0.88($V - J$) + 0.69. This will be further explored in Section \ref{sect:results_uvj_sel}. To summarise, we require:
\begin{itemize} 
\setlength\itemsep{0.1em}
\item $\chi^2_\nu < 6.8$ for GOODS-S or $\chi^2_\nu < 5.3$ for UDS
\item $z_{50} > 2$
\item $M_{*50}  > 10^{10}\mathrm{M_\odot}$
\item sSFR$_{50}$  < 0.2/$t_\mathrm{H}$.
\end{itemize}
\noindent At this point we visually checked the imaging data and posterior distributions for all remaining objects. We identified 21 objects as having very high posterior median redshifts ($5.4 < z < 8.3$). All of these objects have secondary lower-redshift dusty star-forming solutions, and several are required to be extremely massive at their posterior median redshift ($M_*\gtrsim10^{11.5}\mathrm{M_\odot}$). We therefore remove these 21 objects as probable low-redshift dusty interlopers. We also identify 10 poorly fitted objects that are not removed by our $\chi^2_\nu$ criterion (e.g. due to individual strongly discrepant photometry points), which are also rejected. The resulting parent sample of 151 objects is provided in Table \ref{table:sample}. 

\subsection{Selection of a robust sub-sample}\label{sect:method_selection_robust}

It should be stressed that inclusion in our parent sample, shown in Table \ref{table:sample}, does not constitute strong evidence that these objects are $z>2$ quiescent galaxies. Whilst the criteria laid out in Section \ref{sect:method_selection} require that $>50$ per cent of the posterior is consistent with the object being a massive quiescent galaxy at $z>2$, it is still possible that objects have secondary low-redshift or star-forming solutions. We therefore define a further ``robust" sub-sample, for which we can confidently exclude such secondary solutions.

Objects in our robust sub-sample are required to have > 97.5 per cent of their sSFR posterior below sSFR = 0.2/$t_\mathrm{H}$ and < 2.5 per cent of their redshift posterior below $z = 2$. These criteria correspond to 2$\sigma$ upper and lower limits respectively for a Gaussian posterior distribution. To summarise, in addition to the criteria laid down in Section \ref{sect:method_selection}, objects in our robust sub-sample must have:
\begin{itemize}
\setlength\itemsep{0.25em}
\item $z_{2.5} > 2$
\item sSFR$_{97.5}$  < 0.2/$t_\mathrm{H}$.
\end{itemize}
\noindent The 61 resulting objects are flagged as ``robust'' in Table \ref{table:sample}. 

\begin{figure*}
	\includegraphics[width=\textwidth]{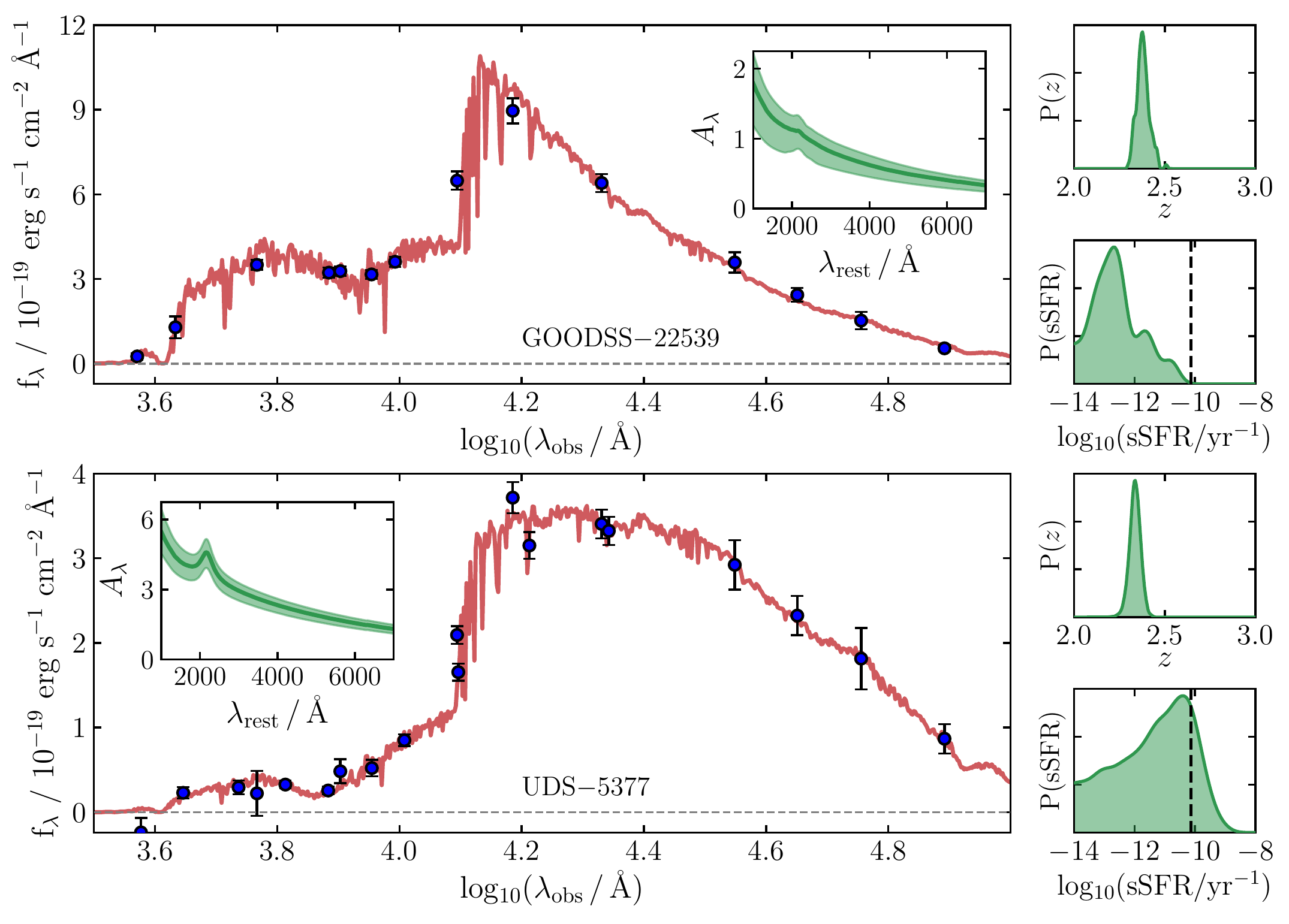}
    \caption{Example objects from our parent sample that do (top) and do not (bottom) meet our robust criteria (see Section \ref{sect:method_selection_robust}). Data are shown in blue, with the posterior median model shown in red. Posterior distributions for redshift, sSFR and dust curve shape are shown in green in the side and inset panels. The dashed vertical lines show our quiescent selection criterion. For the top galaxy $>97.5$ per cent of the sSFR posterior falls below the limit, meaning this object is part of our robust sub-sample. For the bottom object $>50$ per cent of the sSFR posterior falls below the limit, however $>2.5$ per cent falls above. The sSFR of the bottom object is less well constrained as it is fainter and dustier. The sSFR posteriors have tails out to lower values driven by our SFH priors, which are not shown.}\label{fig:example}
\end{figure*}

Fig. \ref{fig:example} shows two example objects from our parent sample that do (top) and do not (bottom) meet our robust selection criteria. Both are at $z\sim2.3{-}2.4$, however the sSFR posterior in the case of the upper object can be seen to be confined more strictly to values below our chosen threshold, which is marked with a dashed black vertical line. The sSFR posterior is better constrained in the case of the upper object firstly because it is brighter and hence detected at higher SNR, and secondly because the lower object is dustier, permitting greater uncertainty in the level of ongoing star formation. In addition, the deeper $H$, $K$ and IRAC data in GOODS-S allows us to place stronger constraints on physical parameters, weighting our robust sub-sample slightly in favour of objects in the GOODS-S field.

\begin{figure*}
	\includegraphics[width=\textwidth]{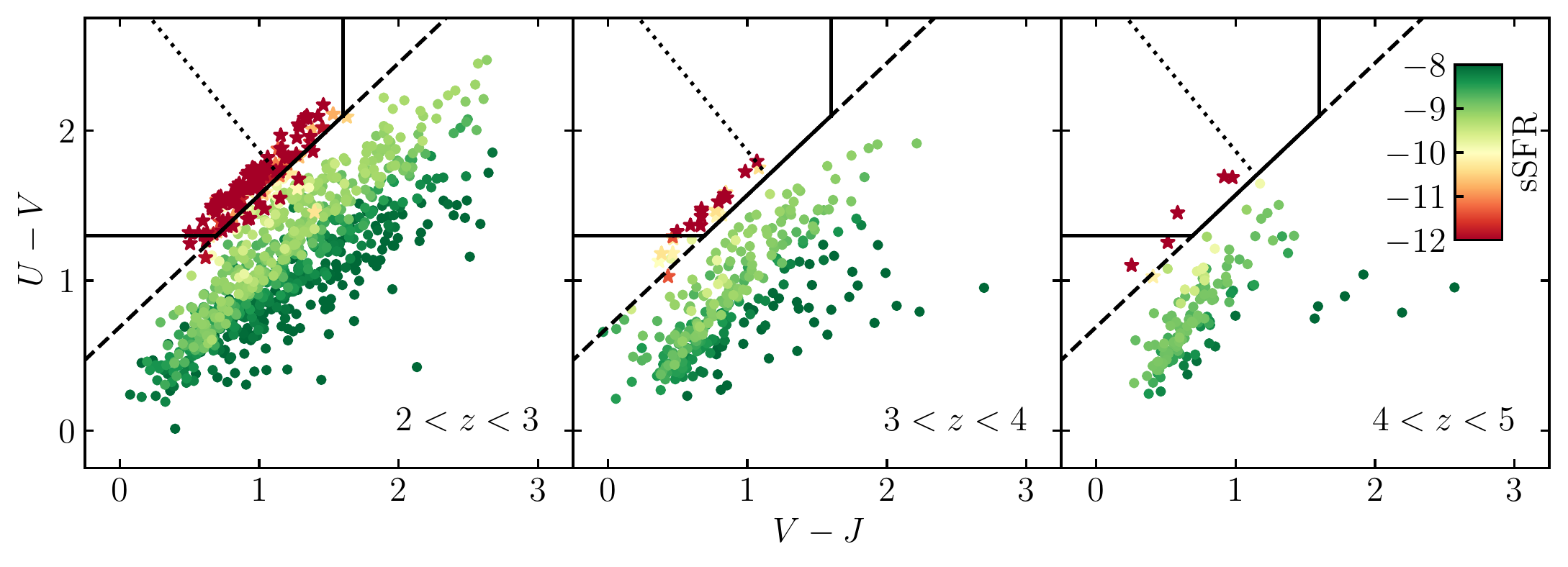}
    \caption{ UVJ diagrams showing the distribution of massive galaxies ($M_* > 10^{10}\mathrm{M_\odot}$) at $2 < z < 5$, coloured by sSFR. Our parent quiescent sample is shown with stars, whilst star-forming galaxies that meet the criteria of Section \ref{sect:method_selection} excepting the sSFR cut are shown with circles. The standard $z=0$ UVJ selection criteria of \protect \cite{Williams2009} are shown with solid black lines, the dashed black line shows the continuation of the diagonal beyond the vertical and horizontal cuts. The dotted line (Equation \ref{eqn:bisector}) bisects the solid diagonal line. At $2 < z < 3$, 25 per cent of our parent sample falls above the bisector, whereas at $z > 3$ only 1 of 41 objects falls above this line.}\label{fig:uvj}
\end{figure*}

\section{Results}\label{sect:results}

\subsection{Quiescent selection at z > 2: UVJ vs sSFR}\label{sect:results_uvj_sel}

Our parent sample is shown on the UVJ diagram in Fig.~\ref{fig:uvj}, split into three redshift bins. Star-forming galaxies that obey the criteria of Section \ref{sect:method_selection} excepting the sSFR threshold are also shown. The standard $z=0$ UVJ selection box of \cite{Williams2009} is shown with black solid lines. The dashed black lines carry on the diagonal beyond the commonly applied vertical and horizontal colour cuts. Following \cite{Williams2009}, we have adopted the $U$ and $V$-band filter curves of \cite{Bessell1990}, and the $J$-band filter curve of \cite{Tokunaga2002}. The median uncertainty on the derived rest-frame colours shown in Fig.~\ref{fig:uvj} is 0.05 magnitudes in $U-V$ and 0.09 magnitudes in $V-J$.

As in \cite{Carnall2018, Carnall2019b}, good agreement can be seen between the \cite{Williams2009} $z=0$ UVJ selection box and the evolving sSFR criterion we have applied. 121 of the 151 galaxies in our (sSFR-selected) parent sample fall within the solid boxes in Fig. \ref{fig:uvj}, along with 1 additional object we identify as star-forming. This suggests the solid box identifies an 80 per cent complete sample of quiescent galaxies at $2 < z < 5$, with a very low contamination rate. 

By contrast, $\sim90$ per cent agreement was found between these selection criteria at $0.25 < z < 2$ by \cite{Carnall2018}. This difference is due to an increased fraction of objects at $z>2$ above the diagonal but below the horizontal cut ($U - V > 1.3$) when compared to lower-redshift samples (10 of 151 objects; 7 per cent), in agreement with the results of \cite{Merlin2018}. This region is strongly associated with post-starburst galaxies at lower redshifts (e.g. \citealt{Wild2014}). When the horizontal colour cut is removed we recover a similar level of agreement to that observed at $z < 2$ (131 of 151 objects; 87 per cent), whilst admitting 3 additional objects we identify as star-forming contaminants.

Extending the solid UVJ selection box in Fig. \ref{fig:uvj} by relaxing the diagonal colour cut to the $z>1$ criterion proposed by \cite{Williams2009}  of $U-V > 0.88(V-J) + 0.49$ admits an additional 15 objects in our parent sample (10 per cent), at the cost of including an additional 35 galaxies that do not meet our sSFR criterion. This implied contamination rate of 20 per cent is in agreement with the results of \cite{Schreiber2018} using the same selection criteria.

\begin{figure*}
	\includegraphics[width=\textwidth]{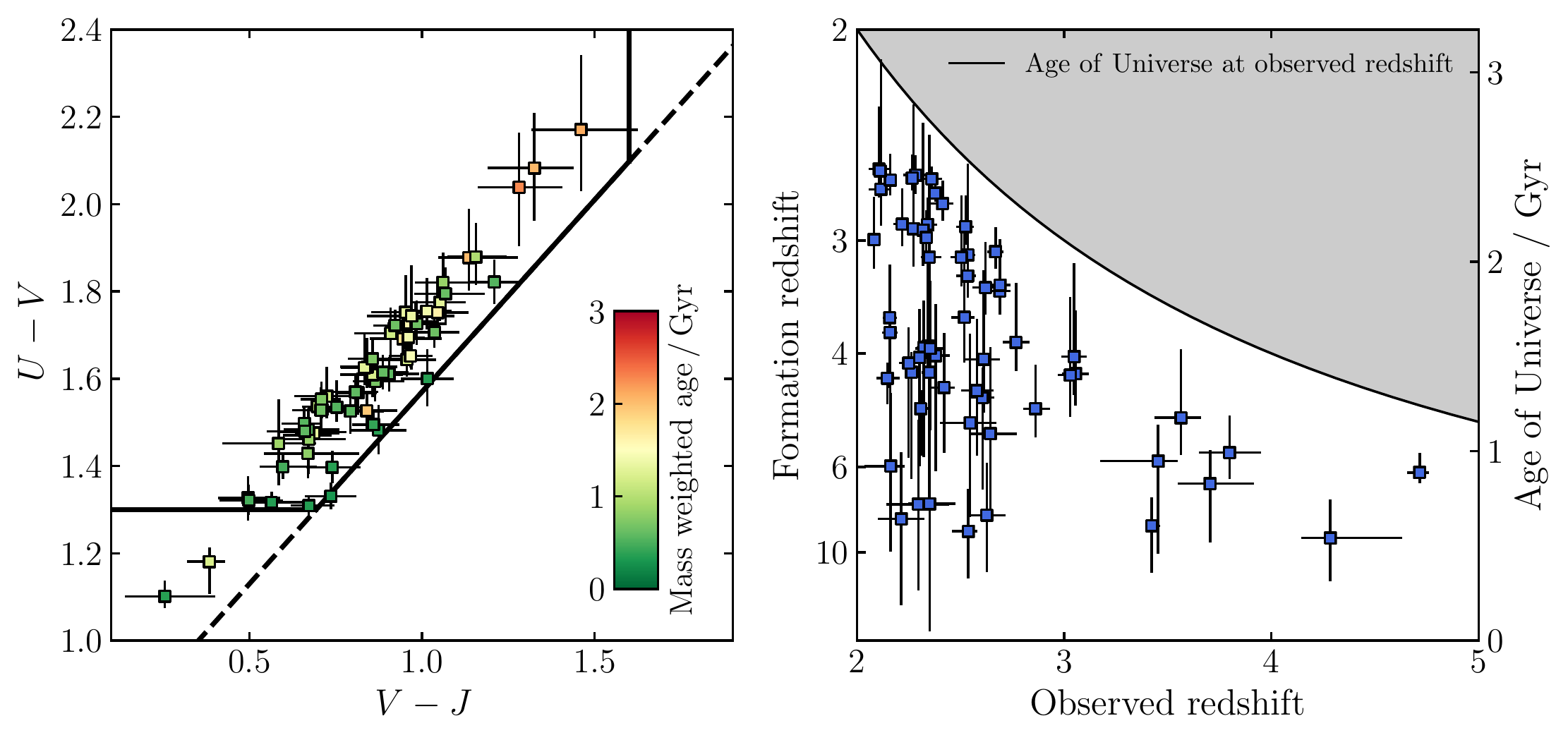}
    \caption{Ages and formation redshifts for galaxies in our robust sub-sample. In the left panel, galaxies are shown on the UVJ diagram coloured by mass-weighted age. The UVJ selection box is as shown in Fig. \ref{fig:uvj}. In the right panel, formation redshifts are shown as a function of observed redshift. A range of formation redshifts are observed, from $z\gtrsim6$ to immediately preceding the observed redshift.}\label{fig:robust}
\end{figure*}

In the left panel of Fig. \ref{fig:robust} we show only our robust sub-sample on the UVJ diagram. It can be seen that 60 of the 61 objects fall above the solid/dashed diagonal selection criterion. All but 2 of these fall within the solid box, with the remaining objects below the horizontal selection criterion.

\subsection{Redshift evolution of quiescent galaxy colours}\label{sect:results_uvj_dist}

As well as playing a role in the separation of the star-forming and quiescent populations, rest-frame UVJ colours are increasingly becoming used as a general tool for understanding galaxy evolution and quenching (e.g. \citealt{Fang2018}; \citealt{Morishita2019}; \citealt{Wu2019}). In particular, \cite{Belli2019} (hereafter \citetalias{Belli2019}) find evidence for a strong age trend within the quiescent population across the UVJ selection box at $1.5 < z < 2.5$, with younger galaxies found towards the bottom-left of the box, and older galaxies found in the upper-right region (see also \citealt{Whitaker2013}). This implies that (assuming no significant redshift evolution in dust attenuation) the upper-right region of the UVJ box should become depopulated at higher redshifts, as not enough time has elapsed since the Big Bang for galaxies to grow old enough to have these colours. 

The left panel of Fig. \ref{fig:uvj} demonstrates (in agreement with e.g. \citealt{Straatman2016}) that galaxies fully populate the diagonal edge of the UVJ selection box by $2 < z < 3$. However, in the central and right panels galaxies appear to be confined to the lower half of the diagonal edge. In order to test whether this difference is significant, we construct the perpendicular bisector for the diagonal edge of the solid UVJ selection box shown in Fig. \ref{fig:uvj}, given by
\begin{equation}\label{eqn:bisector}
U - V = -\dfrac{1}{0.88}(V - J) + 3,
\end{equation}
which is shown with a black dotted line in Fig. \ref{fig:uvj}. 

In the left panel of Fig. \ref{fig:uvj}, at $2 < z < 3$, a total of 27 objects out of 110 in our parent quiescent sample (25 per cent) fall above this bisecting line. If the distribution of colours were the same above $z=3$ we would expect to observe $10\pm3$ objects above this line out of a total of 41. However, as can be seen from Fig. \ref{fig:uvj}, only 1 out of 41 objects in our parent sample at $z>3$ falls above this bisecting line. We therefore find strong evidence that the quiescent population shifts towards bluer colours at $z>3$, consistent with the scheme put forward by \citetalias{Belli2019}.

\subsection{Stellar ages and formation redshifts}\label{sect:results_age}

In this section we report the ages we infer based on our photometric analysis. It should be noted that, as discussed in Section \ref{sect:intro}, age-dating stellar populations based on photometric data is challenging, owing to the age-metallicity-dust degeneracy. 
The flexible model described in Section \ref{sect:method_fitting}, combined with careful exploration of parameter space using \textsc{MultiNest}, allows us to fully characterise the uncertainties on our inferred ages by marginalising over solutions across a wide range of metallicities and dust contents (neither of these properties are strongly constrained by our photometric data). The average uncertainty is $\sim0.5$ Gyr, with the largest uncertainties for the oldest objects.

Age determinations for quiescent galaxies using our double power law SFH model were extensively validated in \cite{Carnall2018}, however no such test has been performed for star-forming galaxies. We therefore limit the discussion in this section to our robust sub-sample. We calculate formation times, $t_\mathrm{form}$, as the average time at which the stars in each galaxy formed,
\begin{equation}
t_\mathrm{form} = \dfrac{\int^{t_\mathrm{obs}}_0 t\ \mathrm{SFR}(t)\ \mathrm{d}t}{\int^{t_\mathrm{obs}}_0 \mathrm{SFR}(t)\ \mathrm{d}t}.
\end{equation}
\noindent This corresponds to the mass-weighted age of the galaxy. We then calculate the redshift corresponding to $t_\mathrm{form}$, which we call the formation redshift. In the left panel of Fig. \ref{fig:robust} we show the mass-weighted ages of galaxies in our robust sub-sample as a function of position on the UVJ diagram. In accordance with the results of Section \ref{sect:results_uvj_sel} it can be seen that we recover a broadly similar age trend across the UVJ box to \citetalias{Belli2019}, albeit with large scatter due to the significant uncertainties on our photometric age determinations.

The right panel of Fig. \ref{fig:robust} shows the formation redshifts we infer as a function of observed redshift. It can be seen that our galaxies exhibit a diverse range of SFHs, with formation redshifts ranging from $z\sim6{-}10$ to immediately preceding the observed redshift. At $z < 3$ the oldest objects are already $\gtrsim2$ Gyr old, making it challenging to set upper limits on their formation redshifts. At $z > 3$ objects are younger and their ages are hence better constrained, with posterior median formation redshifts ranging from $5 < z < 9$. We will further address the question of when the oldest galaxies quenched their star formation in Section \ref{sect:discussion:oldest}.

\subsection{Number density}\label{sect:results_numbers}

The number density of quiescent galaxies at high redshift provides an important constraint on quenching models in simulations of galaxy formation (e.g. \citealt{Dave2017}; \citealt{Merlin2019}; \citealt{Cecchi2019}). In this section we report number densities for our parent sample, as well as our robust sub-sample, which serve as reliable lower limits. We only consider objects at $z < 4$, for which our stellar mass limit of $10^{10}\mathrm{M_\odot}$ constitutes a conservative mass-completeness limit in both fields (e.g. \citealt{Mortlock2015}; McLeod et al. in preparation). The number densities we calculate are given in Table \ref{table:density}, along with uncertainties estimated as the Poisson uncertainty on the number of objects.

\cite{Schreiber2018} report a number density of $2.0\pm0.3\times10^{-5}$ Mpc$^{-3}$ for a spectroscopic sample composed of quiescent galaxies with $K_s < 24.5$ at $3 < z < 4$. In our parent sample 20 galaxies meet these criteria, and we hence derive a number density of $1.7\pm0.4\times10^{-5}$ Mpc$^{-3}$, in good agreement with their result. This suggests that our photometrically selected sample is not strongly contaminated.

\begin{table}
  \caption{Number counts and number densities for objects in our sample ($M_* > 10^{10}\mathrm{M_\odot}$) from $2< z < 4$. The r subscripts denote the robust sub-sample selected in Section \ref{sect:method_selection_robust}, as opposed to the parent sample defined in Section \ref{sect:method_selection}.}
\begin{center}
\begingroup
\setlength{\tabcolsep}{5pt}
\renewcommand{\arraystretch}{1.4}
\begin{tabular}{lcccc}
\hline
Redshift & $N$ & $N_\mathrm{r}$ & $n$ / Mpc$^{-3}$ & $n_\mathrm{r}$ / Mpc$^{-3}$ \\
\hline
$2.0 < z < 2.5$ & 85 & 33 & $1.4\pm0.2\times10^{-4}$ & $5.3\pm0.9\times10^{-5}$ \\
$2.5 < z < 3.0$ & 38 & 18 & $6.2\pm1.0\times10^{-5}$ & $2.9\pm0.7\times10^{-5}$ \\
$3.0 < z < 4.0$ & 22 & 8 & $1.9\pm0.4\times10^{-5}$ & $6.8\pm2.4\times10^{-6}$ \\

\hline
\end{tabular}
\endgroup
\end{center}
\label{table:density}
\end{table}

\section{When did the first galaxies quench?}\label{sect:discussion:oldest}

As discussed in Section \ref{sect:results_age}, the individual ages we derive from our photometric data are highly uncertain. In this section we therefore attempt to use a subset of the most extreme objects in our robust sub-sample to jointly constrain the formation and quenching times of the first galaxies in the Universe to cease star formation.

As demonstrated in Section \ref{sect:results_uvj_dist}, the UVJ colours of the reddest quiescent galaxies evolve with redshift across the range from $2 < z < 5$, becoming bluer at higher redshift as the cosmic time available for them to passively age reduces. We hence select the reddest objects in our robust sub-sample (by $U - V$ colour) in each of the redshift bins shown in Fig.\,\ref{fig:uvj} and attempt to construct a model SFH that traces these objects back to a common redshift of formation.

\begin{figure}
	\includegraphics[width=\columnwidth]{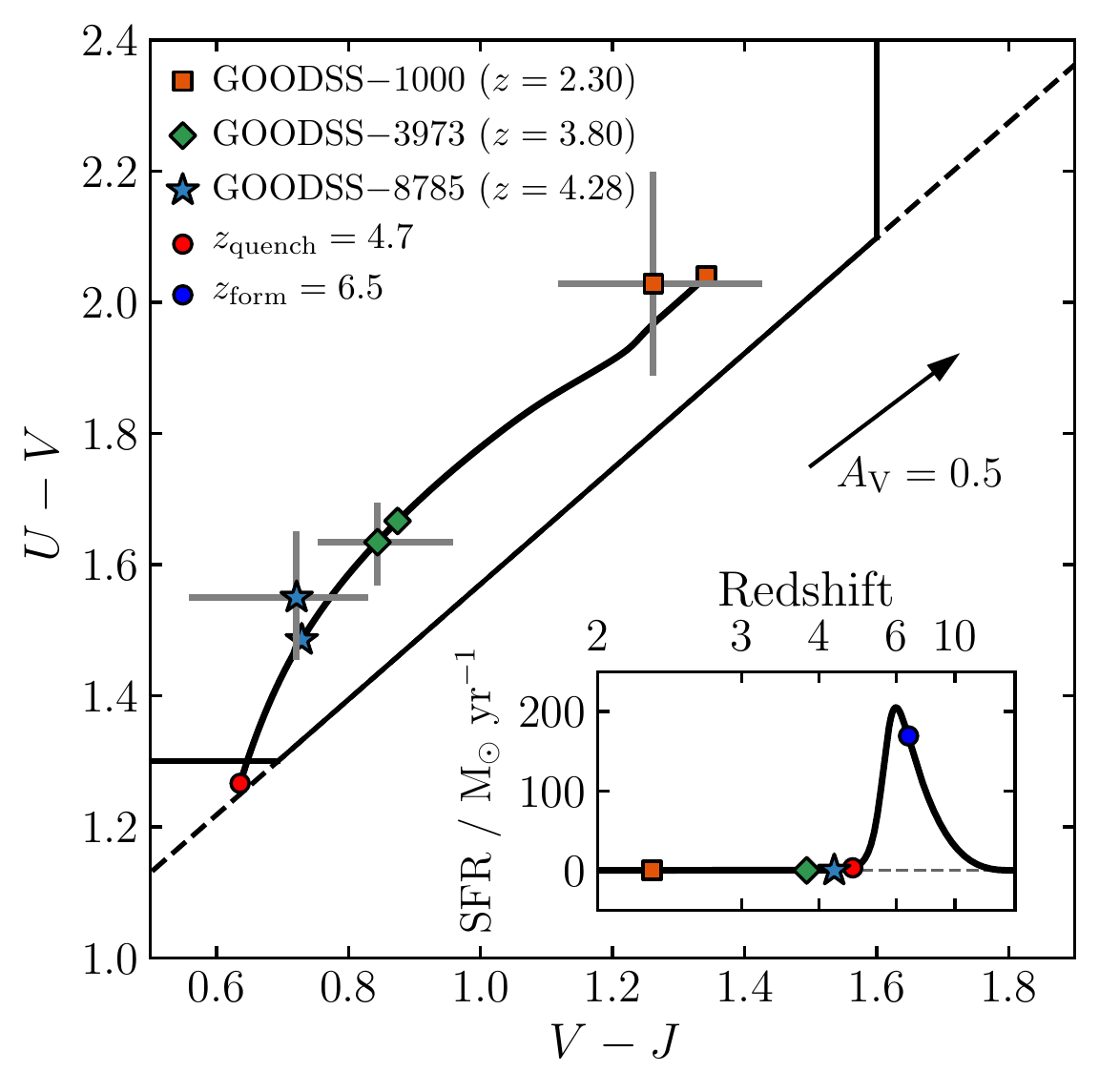}
    \caption{A model for the time-evolution of the UVJ colours of the reddest galaxies in our robust sub-sample in integer redshift bins. These objects, corrected to a common $A_V = 0.58$, are shown as symbols with gray errorbars. The black curved line shows a model for the time-evolution of the UVJ colours of these objects. The corresponding SFH is shown in the inset panel. Symbols superimposed on the line show predicted UVJ colours at the appropriate redshifts. The UVJ selection box is as shown in Fig.\,\ref{fig:uvj}. The \protect\cite{Calzetti2000} dust reddening vector is also shown.}\label{fig:oldest}
\end{figure}

The three objects selected are shown on the UVJ diagram in Fig. \ref{fig:oldest}. As these objects all have different levels of dust attenuation, their positions have been corrected to a common $A_V$ of 0.58, which is the average value for our robust sub-sample, using the \cite{Calzetti2000} law. This average attenuation is higher than is typically found at low redshift, in agreement with other recent analyses (e.g. \citealt{Gobat2018}). The dust attenuation curve slope has little effect on the magnitude of the dust vector shown in Fig. \ref{fig:oldest}, instead resulting in a rotation, which moves objects perpendicular to the diagonal edge of the UVJ box. As we are mainly interested in the evolution of galaxy colours along the age sequence parallel to the diagonal edge of the UVJ box, we do not consider variations in dust attenuation curve shape in this section.

For our model we assume the same constant $A_V$ of 0.58 and further assume fixed Solar metallicity. The model forms a total stellar mass of $10^{11}\mathrm{M_\odot}$, although this has no effect on the resulting UVJ colours. We then vary the SFH shape to attempt to match the UVJ colours of the objects shown in Fig. \ref{fig:oldest} at the appropriate redshifts. 

Our best model is also shown in Fig. \ref{fig:oldest}. The SFH is shown in the inset panel, whilst the black curved line shows the time-evolution of the associated UVJ colours. The symbols plotted along this line correspond to predictions for the UVJ colours of the three objects shown at their observed redshifts. The model SFH has a formation redshift (see Section \ref{sect:results_age}) of $z=6.5$. The model meets our quiescent sSFR selection criterion (see Section \ref{sect:method_selection}) at $z=4.7$. This point is marked with a red circle in Fig. \ref{fig:oldest}. The formation redshifts found through our SED fitting analysis for the objects shown are also consistent with this model.

From this analysis we conclude that the oldest objects in our parent sample are consistent with having formed the bulk of their stellar mass at $z\sim6-7$ and quenched at $z\sim5$.  Earlier formation and quenching is not required to explain our data, however neither is this possibility ruled out. As a final comment, it should be noted that the CANDELS data covers relatively small areas, and it is a clear possibility that more extreme objects may be found in larger area surveys.

\begin{figure*}
	\includegraphics[width=\textwidth]{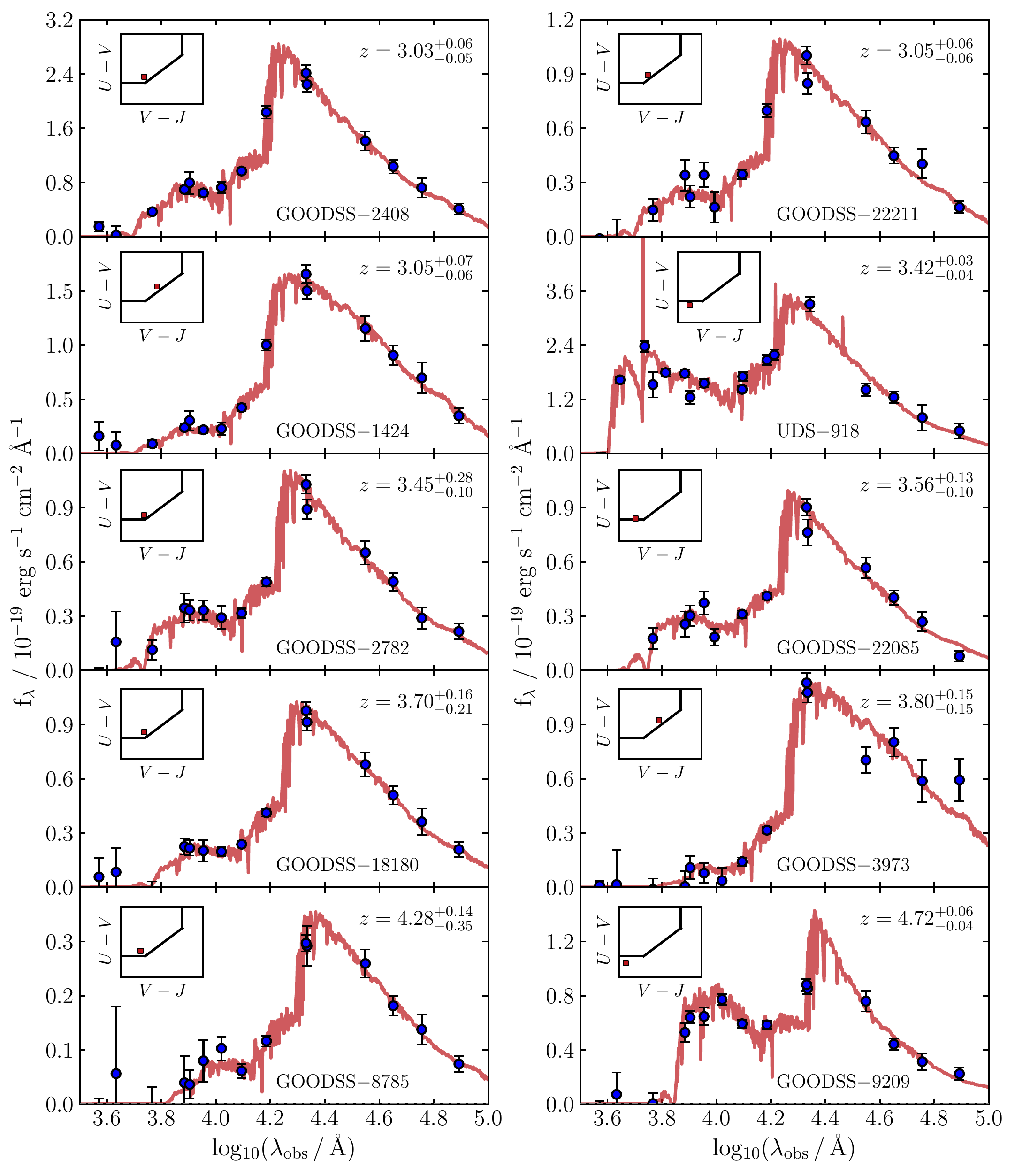}
    \caption{Fitted SEDs for the 10 objects in our robust sub-sample at $z>3$ in ascending order of redshift. Photometric data are shown in blue, whereas our posterior median fitted model is shown in red at our posterior median fitted redshift. The inset panels show the positions of each object on the UVJ diagram, with the solid black lines the same as those shown in Fig. \ref{fig:uvj}. The object GOODSS-2782 was observed for 80 hours as part of VANDELS. The resulting spectrum is shown in the top panel of Fig. \ref{fig:vandels} and is discussed in Section \ref{sect:discussion:2782}. The highest redshift object, GOODSS-9209, was observed for 14 hours as part of VUDS. The resulting spectrum is shown in Fig.\,\ref{fig:vuds} and is disccussed in Section \ref{sect:discussion:9209}.}\label{fig:z3_robust}
\end{figure*}

\begin{figure*}
	\includegraphics[width=0.9\textwidth]{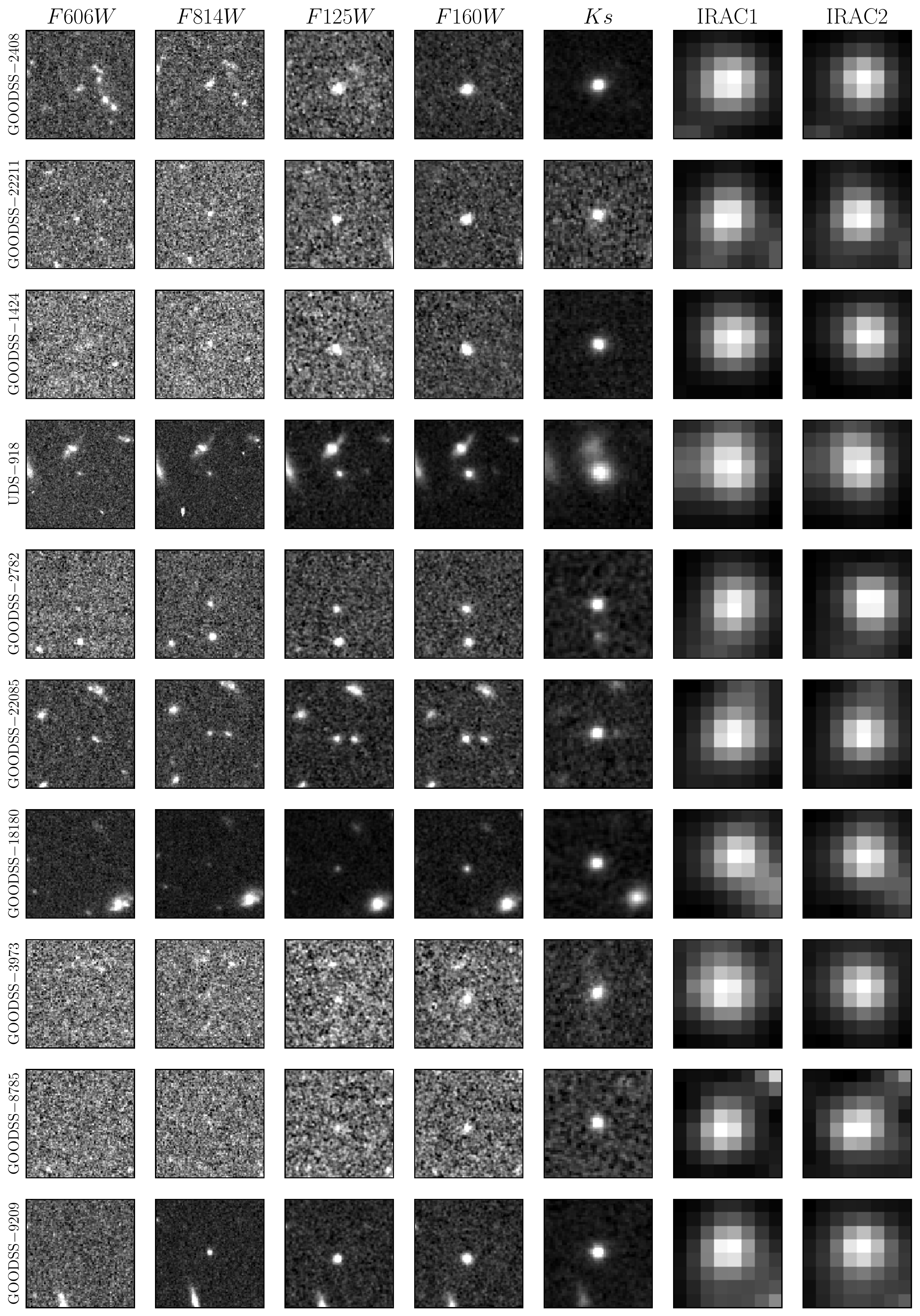}
    \caption{$5''\times5''$ cutouts for each of our 10 robust objects at $z>3$. HST data were taken from CANDELS and Hubble Legacy Field mosaics (\citealt{Teplitz2013, Rafelski2015, Whitaker2019}). IRAC data is from SEDS and S-CANDELS (\citealt{Ashby2015}).}\label{fig:z3_cutouts}
\end{figure*}

\begin{figure*}
	\includegraphics[width=\textwidth]{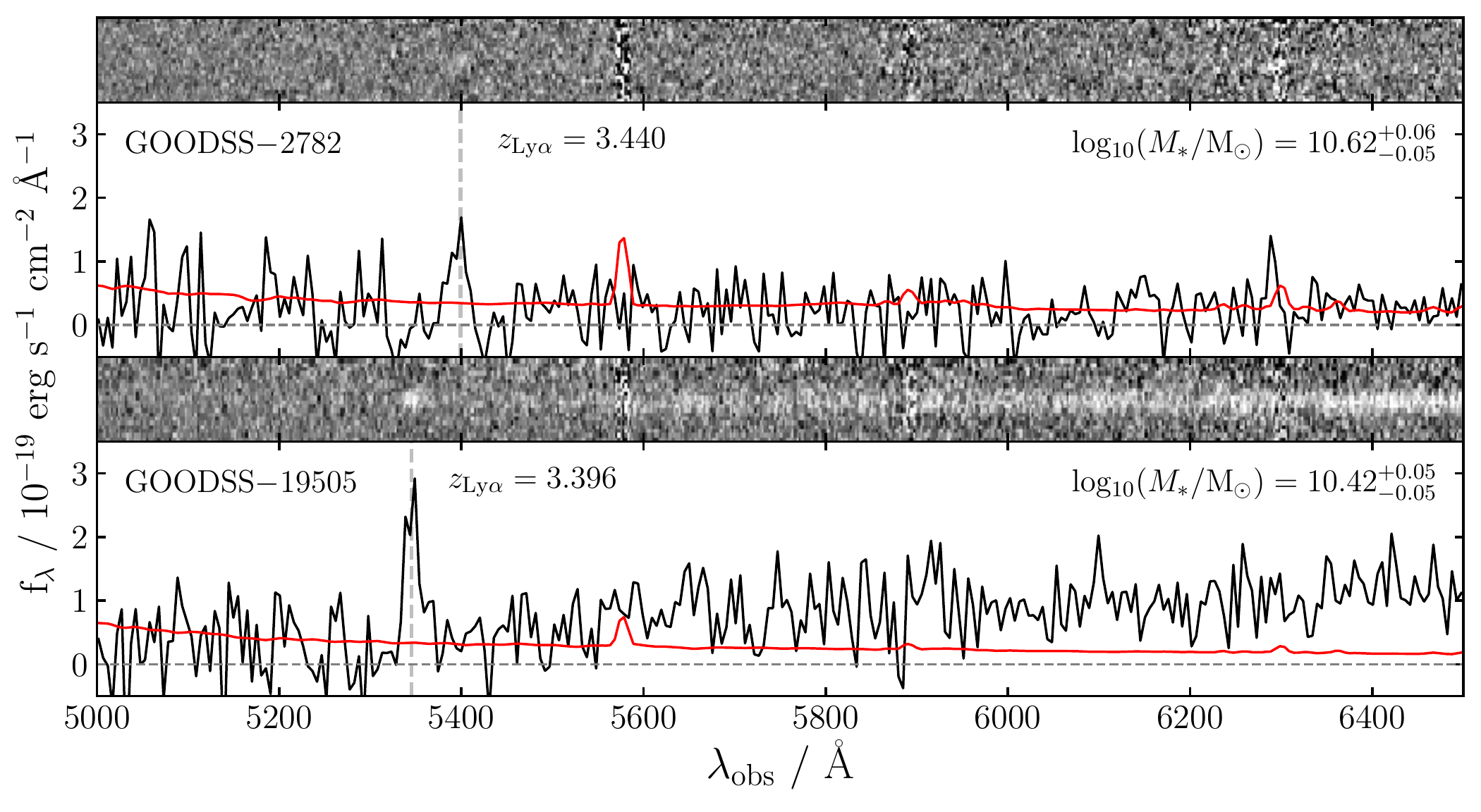}
    \caption{VANDELS spectra for two $z\sim3.4$ galaxies from our sample. The 1D spectra are shown in black with error spectra shown in red. The 2D spectra are shown at the top of each panel. The upper and lower objects received 80 and 40 hours of integration respectively. The upper object is a member of our robust sub-sample, the lower is not. It can be seen that the lower object exhibits stronger Ly$\alpha$ despite its lower mass, implying that our robust selection criteria have been successful in identifying the objects with the lowest sSFRs.}\label{fig:vandels}
\end{figure*}

\section{Massive quiescent galaxies at z > 3}\label{sect:discussion:z3}

Our robust sub-sample contains 10 objects at $z>3$, including 2 at $z > 4$, both of which are at higher redshift than the most distant spectroscopically confirmed quiescent galaxies. As demonstrated in Section \ref{sect:results_age}, the younger ages of quiescent galaxies at $z>3$ make them ideal for placing strong constraints on quenching physics at the highest redshifts. Given this key role in our understanding of galaxy formation, spectroscopic follow-up of such objects is clearly of significant interest. 

In this section, we focus on our robust sub-sample at $z>3$. We firstly compare our results with samples from the literature and then discuss currently available spectroscopic data. The faintness of these objects means that these spectroscopic data are somewhat sparse and of low SNR. However with the next generation of telescopes we can aspire to place strong constraints on the physical properties of such galaxies through high-SNR spectroscopic observations. 

Fitted SEDs for our 10 robust $z > 3$ objects are shown in Fig. \ref{fig:z3_robust}. HST, $K_s$-band and IRAC imaging data for each object are shown in Fig. \ref{fig:z3_cutouts}. These objects can be seen to be generally compact and isolated in the HST imaging.

\subsection{Comparisons with recent studies}

Comparing with other recent work, the $z>3$ robust sub-sample shown in Fig. \ref{fig:z3_robust} contains 6 of the 10 $z > 3$ quiescent  galaxy candidates reported by \cite{Merlin2018} in GOODS-S, with a further 3 of their objects included in our parent sample (which contains 17 objects at $z > 3$ in GOODS-S). The $z > 3$ robust object we identify in UDS is not part of the sample reported by \cite{Merlin2019}, however 7 of their 16 UDS candidates feature in our parent sample (which contains 11 objects at $z > 3$ in UDS). 

\cite{Santini2019} report an analysis of archival ALMA data for the sample of \cite{Merlin2018}, aiming to confirm the passive nature of these objects by placing upper limits on their SFRs. They report non-detections for 3 of our robust $z>3$ objects (GOODSS-2782, 8785 and 9209) and $<3\sigma$ detections for a further two (GOODSS-3973 and 18180). The implied sSFRs for the two ALMA-detected objects are above our selection threshold, however they are consistent to within $2\sigma$ and $1\sigma$ respectively. 

The single UDS $z>3$ object in our robust sub-sample was not observed by \cite{Schreiber2018}, who studied $3 < z < 4$ massive quiescent galaxies in the CANDELS UDS, COSMOS and EGS fields with Keck-MOSFIRE. However three $z>3$ objects from our parent sample were observed. The authors report a spectroscopic redshift of $z=3.543$ for UDS-8682, as well as reporting sSFRs consistent with our quiescent selection criterion for all 3 objects.

\subsection{Spectroscopic observations}

We finally search the many available spectroscopic surveys within UDS and GOODS-S for observations of our robust $z>3$ objects. We find 2 spectra, 1 of which is from VANDELS (\citealt{McLure2018b}; \citealt{Pentericci2018}), and 1 of which is from VUDS (\citealt{LeFevre2015}; \citealt{Tasca2017}). These objects will be discussed in the following sections. We will also discuss a further $z>3$ object from our parent sample that was observed as part of VANDELS.

\begin{figure*}
	\includegraphics[width=\textwidth]{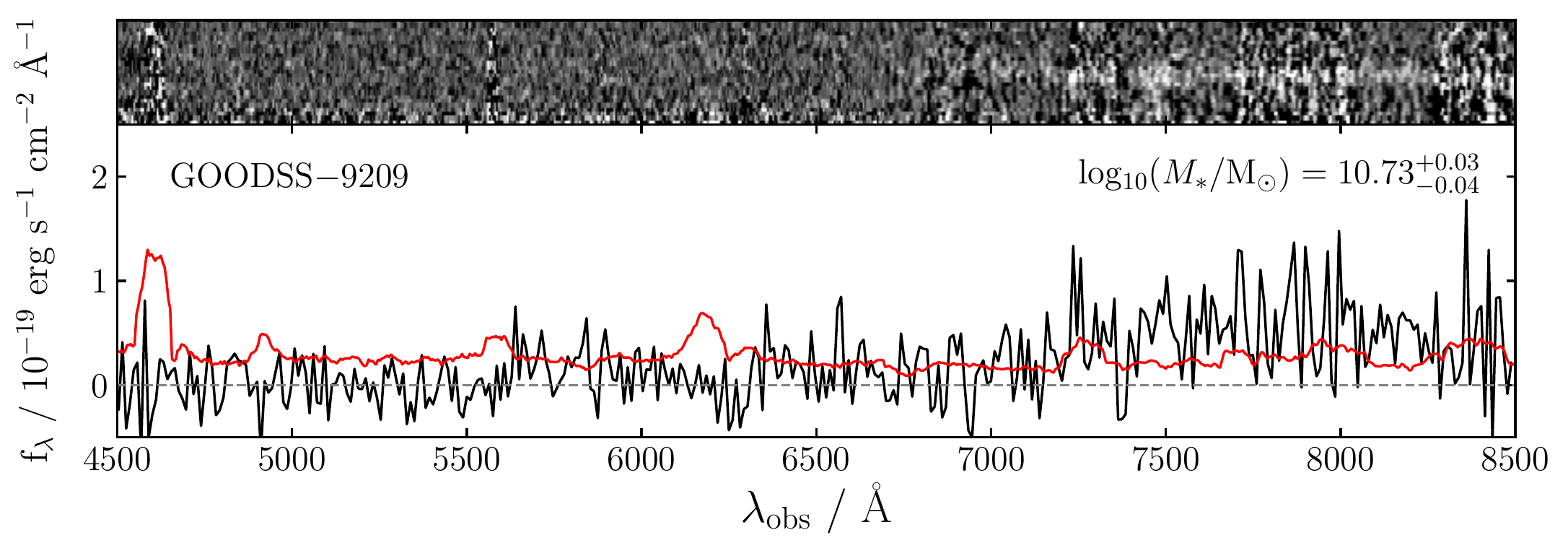}
    \caption{VUDS spectrum for GOODSS-9209, the highest redshift object in our robust sub-sample. The 1D spectrum is shown in black with the error spectrum shown in red. The 2D spectrum is shown at the top. This object received 14 hours of integration. The continuum break in the spectrum at $\lambda\sim7000$\AA\ implies a redshift of $z\sim4.75$, in good agreement with our photometric redshift of $z=4.72^{+0.06}_{-0.04}$.}\label{fig:vuds}
\end{figure*}

\subsubsection{GOODSS-2782}\label{sect:discussion:2782}

VANDELS is a uniquely deep ESO Public spectroscopic survey targeting 2106 high-redshift galaxies with 20, 40 and 80 hour integrations using the VIMOS instrument, providing wavelength coverage from $\sim5000{-}10000$\AA\ at spectral resolution, $R\sim600$. Whilst VANDELS nominally targets star-forming galaxies at $z>2.4$, the chosen criterion of sSFR > 0.1 Gyr$^{-1}$ permits some overlap with our sample at $z>3$. 

GOODSS-2782 is a member of our robust sub-sample, and was targeted by VANDELS for 80 hours of integration. The 1D and 2D spectra for this object are shown in the top panel of Fig. \ref{fig:vandels}. A marginal detection of an emission line can be seen at 5400\AA. For this object we recover a photometric redshift of $z=3.45^{+0.28}_{-0.10}$. This, combined with a CANDELS photometric redshift of $z_\mathrm{CANDELS} = 3.47$, strongly implies this detection is weak Lyman alpha emission at $z=3.440$.

Under this assumption we measure the Ly$\alpha$ flux by first summing the pixels in the 1D spectrum within $\pm30$\AA\ of 5400\AA. We then correct this for continuum emission by subtracting the summed flux of pixels in the 60\AA\ region directly red-wards ($5430{-}5490$\AA). The raw flux within the Ly$\alpha$ region is $2.9\pm0.6\times10^{-18}$ erg s$^{-1}$ cm$^{-2}$, whereas the corrected line flux is $2.1\pm0.9\times10^{-18}$ erg s$^{-1}$ cm$^{-2}$.

It is possible to turn this line flux into an approximation of the SFR for this galaxy, assuming an intrinsic Ly$\alpha$/H$\alpha$ ratio of 8.7 (from case B recombination theory; see \citealt{Osterbrock1989}) and the \cite{Kennicutt2012} relationship between H$\alpha$ luminosity and SFR. This calculation produces a SFR of 0.14$\pm0.07$/$f_\mathrm{esc,\ Ly\alpha}$ M$_\odot$ yr$^{-1}$, where $f_\mathrm{esc,\ Ly\alpha}$ is the escape fraction for Ly$\alpha$. Dividing by the stellar mass posterior yields log$_{10}$(sSFR/yr$^{-1}) = -11.5^{+0.3}_{-0.2} - \mathrm{log_{10}}(f_\mathrm{esc,\,Ly\alpha})$. The sSFR threshold for inclusion in our quiescent sample at $z=3.44$ is log$_{10}$(sSFR/yr$^{-1}) < -9.95$, meaning this object meets our quiescent selection criterion provided that $f_\mathrm{esc,\,Ly\alpha} > 0.03$. Contemporary studies find average Ly$\alpha$ escape fractions of $\sim5-10$ per cent in this redshift range (e.g. \citealt{Sobral2017, Sobral2018}), implying that the strength of Ly$\alpha$ in this spectrum is consistent with this galaxy meeting our sSFR selection criterion.

\cite{Santini2019} report that this galaxy is undetected in ALMA band 7 observations, with an observed flux of $0.04\pm0.11$ mJy/beam, implying a 1$\sigma$ upper limiting SFR of 22.5 M$_\odot$ yr$^{-1}$. To put this into context, a galaxy of this redshift and stellar mass on the star-forming main sequence (SFMS) would have SFR $\sim250$ M$_\odot$ yr$^{-1}$ (\citealt{Speagle2014}), suggesting this galaxy has a SFR suppressed by a factor of $\gtrsim11$ relative to the SFMS.

\subsubsection{GOODSS-19505}\label{sect:discussion:19505}

GOODSS-19505 is a member of our parent sample but not a member of our robust sub-sample, with 28 per cent of the \bagpipes\ sSFR posterior above the 0.2/$t_\mathrm{H}$ threshold. This object was targeted by VANDELS for 40 hours of integration. The resulting spectrum is shown in the bottom panel of Fig. \ref{fig:vandels}. This spectrum shows a much clearer emission line, this time at 5350\AA, as well as continuum emission at longer wavelengths. The photometric redshift we recover of $z=3.49^{+0.09}_{-0.27}$ and CANDELS photometric redshift of $z_\mathrm{CANDELS} = 3.47$ are again consistent with the identification of this line as Ly$\alpha$, this time at $z=3.396$.

Following the same procedure as Section \ref{sect:discussion:2782} we find a raw line flux of $6.2\pm0.6\times10^{-18}$ erg s$^{-1}$ cm$^{-2}$ and a corrected line flux of $4.4\pm0.8\times10^{-18}$ erg s$^{-1}$ cm$^{-2}$. This implies a SFR of 0.28$\pm0.05$/$f_\mathrm{esc,\,Ly\alpha}$ M$_\odot$ yr$^{-1}$ and therefore log$_{10}$(sSFR/yr$^{-1}) = -11.0^{+0.2}_{-0.3} - \mathrm{log_{10}}(f_\mathrm{esc,\,Ly\alpha})$. An escape fraction, $f_\mathrm{esc,\,Ly\alpha} > 0.1$ is therefore required for this object to be consistent with our quiescent selection criterion.

Our finding of stronger Ly$\alpha$ and an implied higher sSFR for this object compared to GOODSS-2782 is encouraging, suggesting that our robust selection criteria are correctly identifying objects with the most suppressed sSFRs.

\cite{Santini2019} also report that this galaxy is undetected in ALMA band 6 observations, with an observed flux of $0.02\pm0.4$ mJy/beam, implying a 1$\sigma$ upper limiting SFR of 20.6 M$_\odot$ yr$^{-1}$. A galaxy of this redshift and stellar mass on the SFMS would have SFR $\sim180$ M$_\odot$ yr$^{-1}$, meaning that the SFR this galaxy is suppressed by a factor of $\gtrsim9$ relative to the SFMS.

\subsubsection{GOODSS-9209}\label{sect:discussion:9209}

GOODSS-9209 is the highest redshift object in our robust sub-sample, shown in the bottom-right panel of Fig. \ref{fig:z3_robust} and the bottom row of Fig. \ref{fig:z3_cutouts}. This object was observed for 14 hours as part of VUDS, the resulting 2D and 1D spectra are shown in Fig. \ref{fig:vuds}. The spectrum exhibits a continuum break around $\lambda \sim 7000$\AA, consistent with a redshift of $z\sim4.75$. The VUDS team measure a redshift of $z=4.657$, with a quality flag of 1, indicating 50 per cent probability of correct identification. The spectrum exhibits no obvious Ly$\alpha$ emission, however the redshift implied by the continuum break is consistent with the photometric redshift we derive with \bagpipes, of $z=4.72^{+0.06}_{-0.04}$. If confirmed, this would be the highest redshift known quiescent galaxy. 

\cite{Santini2019} report that this galaxy is undetected in ALMA band 6 observations, with an observed flux of $-0.03\pm0.16$ mJy/beam, implying a 1$\sigma$ upper limiting SFR of 41.6 M$_\odot$ yr$^{-1}$. A galaxy of the same stellar mass on the SFMS at this redshift would have SFR $\sim300$ M$_\odot$ yr$^{-1}$. This implies the SFR of this galaxy is suppressed by a factor of $\gtrsim7.5$ relative to the main sequence.

\section{Conclusion}\label{sect:conclusion}

In this work we have selected a sample of 151 massive ($M_* > 10^{10}\mathrm{M_\odot}$) quiescent galaxies at $2 < z < 5$ from high-quality photometric data in the CANDELS UDS and GOODS-South fields. We apply a sophisticated Bayesian fitting approach, using the \bagpipes\ code (\citealt{Carnall2018}) to fit a nine-parameter model to our data, including freedom in the observed redshift, a flexible dust curve shape and double power law SFH model. This approach is designed to identify all potential solutions across a wide range of sSFRs and observed redshifts. Our sample includes a robust sub-sample of 61 objects for which we can exclude star-forming and low-redshift solutions with high confidence. Our robust sub-sample includes 10 objects at $z>3$, of which 2 are at $z>4$. A comparison of number densities between our sample and the spectroscopic sample of \cite{Schreiber2018} at $3 < z < 4$ yields good agreement, implying that our \bagpipes\ fitting approach has been successful and confirming that our sample is not strongly contaminated.

We report rest-frame UVJ colours for our sample, demonstrating that the distribution of UVJ colours for the quiescent population becomes bluer with increasing redshift at $z\gtrsim3$. This is predicted by the model of \cite{Belli2019}, in which galaxies quench close to the lower left edge of the box, then passively age upwards and to the right. In addition, we identify a substantial population of quiescent objects below the standard $U-V=1.3$ colour cut, a region that is sparsely populated at low redshift, and is strongly associated with post-starburst galaxies (e.g. \citealt{Wild2014}).

We report individual formation redshifts for the objects in our robust sub-sample, demonstrating that the $z>2$ quiescent population is diverse, with formation redshifts ranging from $z\sim6{-}10$ to immediately preceding the redshift of observation. However, the individual ages and SFHs we obtain from our photometric data are significantly uncertain, particularly for older objects at $z<3$, with an average uncertainty of $\sim0.5$ Gyr on our inferred formation times.

These findings demonstrate the need for high-SNR continuum spectroscopic data to constrain the formation epochs and maximum historical SFRs of these galaxies (e.g. \citealt{Carnall2019b}). Such analyses would allow us to better constrain high-redshift quenching physics, and to quantify the widely discussed connection between high redshift massive quiescent galaxies and rapidly star-forming submillimetre galaxies (e.g. \citealt{Stach2019}; \citealt{Wild2020}).

In an attempt to more strongly constrain the formation and quenching times of the oldest galaxies from photometry, we select the reddest, and hence oldest, objects in our robust sub-sample in integer redshift bins and construct a model for the time-evolution of their UVJ colours. We find that the UVJ colours of these oldest objects can be explained by a model with a formation redshift of $z=6.5$ and quenching redshift of $z=4.7$. We find no evidence that quenching before $z=5$ is required to explain the stellar populations we observe, though earlier formation is not ruled out by these data.

We report spectroscopic redshifts for two galaxies in our sample at $z\sim3.4$, one of which is a member of our robust sub-sample. These galaxies exhibit weak Ly$\alpha$ emission in ultra-deep, 40 and 80 hour optical spectra from VANDELS (\citealt{McLure2018b}; \citealt{Pentericci2018}). We estimate the SFRs implied by their Ly$\alpha$ fluxes, demonstrating that they are consistent with our quiescent galaxy selection criterion provided that these objects have Ly$\alpha$ escape fractions, $f_\mathrm{esc,\,Ly\alpha} > 0.03$ and $f_\mathrm{esc,\,Ly\alpha} > 0.1$ for the robust and non-robust objects respectively. This suggests our robust selection criteria have been successful in identifying the objects with the lowest sSFRs. These objects are reported as being undetected in ALMA observations by \cite{Santini2019}, implying SFRs suppressed by a factor of $\gtrsim10$ compared to the star-forming main sequence at this redshift.

We finally present a spectrum from VUDS for the highest redshift object in our robust sub-sample, for which we find a photometric redshift of $z=4.72^{+0.06}_{-0.04}$. This object exhibits a continuum break at $\lambda\sim7000$\AA, implying a redshift of $z\sim4.75$, consistent with our photometric redshift. This object is also undetected in ALMA observations, suggesting a SFR suppressed by a factor of $\gtrsim7.5$ compared to the SFMS at our photometric redshift. If spectroscopically confirmed this would be the highest redshift known quiescent galaxy.

\section*{Acknowledgements}

A. C. Carnall and F. Cullen acknowledge the support of the UK Science and Technology Facilities Council. A. Cimatti acknowledges the grants ASI n.2018-23-HH.0, PRIN MIUR 2015 and PRIN MIUR 2017 - 20173ML3WW\_001. This work is based on data products from observations made with ESO Telescopes at La Silla Paranal Observatory under ESO programme ID 194.A-2003(E-Q). This work is based in part on observations made with the Spitzer Space Telescope, which is operated by the Jet Propulsion Laboratory, California Institute of Technology under a NASA contract. This research made use of Astropy, a community-developed core Python package for Astronomy \citep{Astropy2013}.

\bibliographystyle{mnras}
\bibliography{carnall2020a} 

\bsp
\label{lastpage}
\end{document}